\documentstyle[aip,11pt]{article}
\def\baselinestretch{1.4}
\def\barr{\begin{array}}
\def\btab{\begin{tabular}}
\def\etab{\end{tabular}}
\def\bmi{\begin{minipage}}
\def\emi{\end{minipage}}
\def\bbi{\begin{thebibliography}{99}}
\def\ebi{\end{thebibliography}}
\def\bce{\begin{center}}
\def\ece{\end{center}}
\def\bab{\begin{abstract}}
\def\eab{\end{abstract}}
\def\ear{\end{array}}
\def\earr{\end{array}}
\def\beq{\begin{equation}}
\def\beqa{\begin{eqnarray}}
\def\eeqa{\end{eqnarray}}
\def\eeq{\end{equation}}
\def\ben{\begin{enumerate}}
\def\een{\end{enumerate}}
\def\bdo{
\begin{document}}
\def\edo{\end{document}}
\def\bit{\begin{itemize}}
\def\eit{\end{itemize}}
\def\ble{\begin{letter}}
\def\ele{\end{\letter}}
\def\sech{\mbox{sech}}
\def\Tr{\mbox{Tr}}
\def\ol#1{\overline{#1}}
\def\fig#1{\newpage\vspace*{7.4in}\bce Figure #1\ece}
\def\rta{\rightarrow}
\def\lra{\Leftrightarrow}
\def\rlh{\rightleftharpoons}
\def\eg{{\it e.g., }}
\def\ap{{\it a priori }}
\def\cf{{\it cf}.~}
\def\ie{{\it i.e., }}
\def\etal{{\it et al.}}
\def\etc{{\it etc.}}
\def\vs{{\it vs.}~}
\def\via{{\it via} }
\def\hsph{\hspace*{0.5in}}
\def\hspq{\hspace*{0.25in}}
\def\andh{\hspace*{0.25in}\mbox{and}\hspace*{0.25in}}
\def\orh{\hspace*{0.25in}\mbox{or}\hspace*{0.25in}}
\def\wheh{\hspace*{0.25in}\mbox{where}\hspace*{0.25in}}
\def\forh{\hspace*{0.25in}\mbox{for}\hspace*{0.25in}}
\def\dash{\mbox{-}}
\def\vr{\vec{r}}
\def\vR{\vec{R}}
\def\vq{\vec{q}}
\def\vG{\vec{G}}
\def\vA{\vec{A}}
\def\vB{\vec{B}}
\def\vl{\vec{l}}
\def\sd{d\mbox{\put(-3.5,6){\line(1,0){5}}}}
\def\soli{\mbox{(\put(0,3){\line(1,0){49}}\hsp{0.7})}}
\def\dali{\mbox{(\put(0,3){\line(1,0){7}\,\,\line(1,0){7}\,\,\line(1,0){7}
          \,\,\line(1,0){7}\,\,\line(1,0){7}}\hsp{0.7})}}
\def\sdli{\mbox{(\put(0,3){\line(1,0){3}\,\line(1,0){3}\,\line(1,0){3}
          \,\line(1,0){3}\,\line(1,0){3}\,\line(1,0){3}\,\line(1,0){3}
          \,\line(1,0){3}\,\line(1,0){3}\,\line(1,0){3}\,\line(1,0){3}}
          \hsp{0.7})}}
\def\udli{\mbox{(\put(0,3){\line(1,0){5}\,\line(1,0){2}\,\line(1,0){5}
          \,\line(1,0){2}\,\line(1,0){5}\,\line(1,0){2}\,\line(1,0){5}
          \,\line(1,0){2}\,\line(1,0){5}\,\line(1,0){2}}
          \hsp{0.7})}}
\def\doli{\mbox{(\put(0,2.5){.\,.\,.\,.\,.\,.\,.\,.\,.\,.\,.}\hsp{0.7})}}
\def\ddli{\mbox{(\put(0,2.5){\_\,.\,\_\,.\,\_\,.\,\_\,.\,\_\,.}\hsp{0.7})}}
\def\pll{_\parallel}
\def\pllz{_{\parallel0}}
\def\per{_\perp}
\def\perz{_{\perp0}}
\def\al{\alpha}
\def\all{{\;\;\al}}
\def\bel{{\;\;\be}}
\def\gal{{\;\;\ga}}
\def\del{{\;\;\de}}
\def\allu{^{\;\;\al}}
\def\alld{_{\;\;\al}}
\def\belu{^{\;\;\be}}
\def\beld{_{\;\;\be}}
\def\galu{^{\;\;\ga}}
\def\gald{_{\;\;\ga}}
\def\delu{^{\;\;\de}}
\def\deld{_{\;\;\de}}
\def\umi{^{(-)}}
\def\upl{^{(+)}}
\def\upm{^{(\pm)}}
\def\ump{^{(\mp)}}
\def\uze{^{(0)}}
\def\be{{\beta}}
\def\bed{\dot{\beta}}
\def\qd{\dot{q}}
\def\pd{\dot{p}}
\def\Qd{\dot{Q}}
\def\Pd{\dot{P}}
\def\qdd{\ddot{q}}
\def\ga{{\gamma}}
\def\de{{\delta}}
\def\ep{\epsilon}
\def\ze{\zeta}
\def\et{\eta}
\def\th{\theta}
\def\io{\iota}
\def\ka{{\kappa}}
\def\la{\lambda}
\def\om{omicron}
\def\rh{\rho}
\def\si{{\sigma}}
\def\ta{\tau}
\def\up{\upsilon}
\def\ph{\phi}
\def\ch{\chi}
\def\ps{\psi}
\def\om{{\omega}}
\def\prl#1#2#3{, Phys.\ Rev.\ Lett.\ {\bf #1}, #2 (#3)}
\def\prb#1#2#3{, Phys.\ Rev.\ B {\bf #1}, #2 (#3)}
\def\pra#1#2#3{, Phys.\ Rev.\ A {\bf #1}, #2 (#3)}
\def\jcp#1#2#3{, J.\ Chem.\ Phys.\ {\bf #1}, #2 (#3)}
\def\rmp#1#2#3{, Rev.\ Mod.\ Phys.\ {\bf #1}, #2 (#3)}
\def\ssc#1#2#3{, Solid State Commun.~{\bf #1}, #2 (#3)}
\def\mop#1#2#3{, Mol.~Phys.~{\bf #1}, #2 (#3)}
\def\jpc#1#2#3{, J.~Phys.~Chem.~{\bf #1}, #2 (#3)}
\def\cpl#1#2#3{, Chem.~Phys.~Lett.~{\bf #1}, #2 (#3)}
\def\ab{\bar{a}}
\def\bb{\bar{b}}
\def\Tb{\bar{T}}
\def\Jb{\bar{J}}
\def\Kb{\bar{K}}
\def\Sb{\bar{S}}
\def\Ab{\bar{A}}
\def\At{\tilde{A}}
\def\Et{\tilde{E}}
\def\yt{\tilde{y}}
\def\Fb{\bar{F}}
\def\gb{\bar{g}}
\def\xb{\bar{x}}
\def\pb{\bar{p}}
\def\vSb{{\bf \bar{S}}}
\def\vKb{{\bf \bar{K}}}
\def\vTb{{\bf \bar{T}}}
\def\vPb{{\bf \bar{P}}}
\def\vIb{{\bf \bar{I}}}
\def\f{{\bf f}}
\def\ftp{{\bf f}(\th,\ph)}
\def\a{{\bf a}}
\def\at{{\bf a}(t)}
\def\ao{{\bf a}(\om)}
\def\F{{\bf F}}
\def\E{{\bf E}}
\def\B{{\bf B}}
\def\C{{\bf C}}
\def\x{{\bf x}}
\def\xd{{\dot{\bf x}}}
\def\vd{{\dot{\bf v}}}
\def\rdv{{\dot{\bf r}}}
\def\pdv{{\dot{\bf p}}}
\def\rddv{{\ddot{\bf r}}}
\def\rd{\dot r}
\def\ad{\dot a}
\def\add{\ddot a}
\def\thd{\dot\th}
\def\phd{\dot\ph}
\def\rdd{\ddot r}
\def\phdd{\ddot\ph}
\def\thdd{\ddot\th}
\def\xt{{\bf x}(t)}
\def\tx{(t,\x)}
\def\tpxp{(t',\xp)}
\def\txu{(t_1,\x_1)}
\def\txd{(t_2,\x_2)}
\def\k{{\bf k}}
\def\q{{\bf q}}
\def\kp{{\bf k'}}
\def\xp{{\bf x'}}
\def\A{{\bf A}}
\def\H{{\bf H}}
\def\D{{\bf D}}
\def\I{{\bf I}}
\def\K{{\bf K}}
\def\S{{\bf S}}
\def\J{{\bf J}}
\def\j{{\bf j}}
\def\M{{\bf M}}
\def\P{{\bf P}}
\def\V{{\bf V}}
\def\U{{\bf U}}
\def\X{{\bf X}}
\def\R{{\bf R}}
\def\Q{{\bf Q}}
\def\Om{\Omega}
\def\Ph{\Phi}
\def\mn{\mu\nu}
\def\nm{\nu\mu}
\def\CF{{\cal F}}
\def\CFv{\mbox{\boldmath{\cal F}\unboldmath}}
\def\CH{{\cal H}}
\def\CO{{\cal O}}
\def\CC{{\cal C}}
\def\CM{{\cal M}}
\def\CN{{\cal N}}
\def\CP{{\cal P}}
\def\CW{{\cal W}}
\def\CZ{{\cal Z}}
\def\CA{{\cal A}}
\def\ul{\underline}
\def\aboverel#1\above#2{\mathrel{\mathop{#2}\limits^{#1}}}
\def\beneathrel#1\under#2{\mathrel{\mathop{#2}\limits_{#1}}}
\def\frac#1#2{{#1 \over #2}}
\def\vev#1{{\left\langle{#1}\right\rangle}}
\def\state#1{|#1\rangle}
\def\vac{\state0}
\def\sol{\state s}
\def\div{\nabla\cdot}
\def\divp{\nabla'\cdot}
\def\grad{\nabla}
\def\curl{\nabla\times}
\def\curlp{\nabla'\times}
\def\lap{\nabla^2}
\def\dal{\Box^2}
\def\pde#1#2{\frac{\partial #1}{\partial #2}}
\def\pdep#1#2{\left({\frac{\partial #1}{\partial #2}}\right)}
\def\pdes#1#2#3{\lep\frac{\partial #1}{\partial #2}\rip_{#3}}
\def\pop{\partial}
\def\der#1#2{\frac{d #1}{d #2}}
\def\dpdo{\der{{\cal P}}\Om}
\def\spd#1#2{\frac{\partial^2 #1}{\partial #2^2}}
\def\spdm#1#2#3{\frac{\partial^2 #1}{\partial #2\partial #3}}
\def\spds#1#2#3{\lep\frac{\partial^2 #1}{\partial #2^2}\rip_{#3}}
\def\sq{{\vbox {\hrule height 0.6pt\hbox{\vrule width 0.6pt\hskip 3pt
        \vbox{\vskip 6pt}\hskip 3pt \vrule width 0.6pt}\hrule height 0.6pt}}}
\def\square{\kern1pt\vbox{\hrule height 1.2pt\hbox{\vrule width 1.2pt\hskip 3pt
   \vbox{\vskip 6pt}\hskip 3pt\vrule width 0.6pt}\hrule height 0.6pt}\kern1pt}
\def\nn{{\bf n}}
\def\nnz{{\bf n}_0}
\def\np{{\bf n'}}
\def\epb{\bar{\ep}}
\def\thb{\bar{\th}}
\def\scl{{\cal L}}
\def\thu{\mbox{{\boldmath $\th$}}}
\def\phu{\mbox{{\boldmath $\ph$}}}
\def\epi{\mbox{{\boldmath $\ep_i$}}}
\def\epv{\mbox{{\boldmath $\ep$}}}
\def\rhv{\mbox{{\boldmath $\rh$}}}
\def\epz{\mbox{{\boldmath $\ep$}}_0}
\def\epzs{\mbox{{\boldmath $\ep$}}^*_0}
\def\epj{\mbox{{\boldmath $\ep_j$}}}
\def\epk{\mbox{{\boldmath $\ep_k$}}}
\def\epu{\mbox{{\boldmath $\ep_1$}}}
\def\epd{\mbox{{\boldmath $\ep_2$}}}
\def\ept{\mbox{{\boldmath $\ep_3$}}}
\def\epxx{\mbox{{\boldmath $\ep_x$}}}
\def\epyy{\mbox{{\boldmath $\ep_y$}}}
\def\epzz{\mbox{{\boldmath $\ep_z$}}}
\def\eppm{\mbox{{\boldmath $\ep_\pm$}}}
\def\epmp{\mbox{{\boldmath $\ep_\mp$}}}
\def\epp{\mbox{{\boldmath $\ep_+$}}}
\def\epm{\mbox{{\boldmath $\ep_-$}}}
\def\vbed{\dot{\mbox{{\boldmath $\be$}}}}
\def\thh{\hat{\mbox{{\boldmath $\theta$}}}}
\def\phh{\hat{\mbox{{\boldmath $\phi$}}}}
\def\rhoh{\hat{\mbox{{\boldmath $\rho$}}}}
\def\vbe{\mbox{{\boldmath $\be$}}}
\def\vom{\mbox{{\boldmath $\om$}}}
\def\vze{\mbox{{\boldmath $\ze$}}}
\def\vpi{\mbox{{\boldmath $\pi$}}}
\def\vth{\mbox{{\boldmath $\th$}}}
\def\epo{\ep(\om)}
\def\epop{\ep(\om')}
\def\g{{\bf g}}
\def\dV{{\bf\bar{V}}}
\def\dT{{\bf\bar{T}}}
\def\iniv{\int d^3x\,}
\def\inia{\int d^2x\,}
\def\inivp{\int d^3x'\,}
\def\iniap{\int d^2x'\,}
\def\inik{\int d^3k\,}
\def\inxt{\int d^3x\,dt\,}
\def\inxtp{\int d^3x'dt'}
\def\dvp{d^3x'\,}
\def\dap{d^2x'\,}
\def\inv{\int_V d^3x\,}
\def\invp{\int_V d^3x'\,}
\def\ina{\int_S d^2x\,}
\def\inac{\oint_S d^2x\,}
\def\inacp{\oint_Sd^2x'\,}
\def\inap{\int_S d^2x'\,}
\def\invl{\int_C d{\bf l}}
\def\invlc{\oint_C d{\bf l}}
\def\inlc{\oint_C dl}
\def\inopc{\oint_C d\om'\,}
\def\invv{\int d^3x\,d^3x'\,}
\def\inko{\int d^3k\,d\om}
\def\intmp{\int_{-\infty}^\infty}
\def\intzp{\int_0^\infty}
\def\ftk{\frac1{\sqrt{2\pi}}\int_{-\infty}^\infty dk\,}
\def\ftz{\frac1{\sqrt{2\pi}}\int_{-\infty}^\infty dz\,}
\def\fto{\frac1{\sqrt{2\pi}}\int_{-\infty}^\infty d\om\,}
\def\ftt{\frac1{\sqrt{2\pi}}\int_{-\infty}^\infty dt\,}
\def\stp{\sqrt{2\pi}}
\def\fpi{{4\pi}}
\def\tpi{{2\pi}}
\def\tpic{{(\tpi)^3}}
\def\embe{e^{-\be E}}
\def\emot{e^{-i\om t}}
\def\eot{e^{i\om t}}
\def\eotp{e^{i\om t'}}
\def\emopt{e^{-i\om't}}
\def\etot{e^{-2i\om t}}
\def\ekxot{e^{i(\k\cdot\x-\om t)}}
\def\emkxot{e^{-i(\k\cdot\x-\om t)}}
\def\ekzot{e^{i(kz-\om t)}}
\def\ekzokt{e^{i(kz-\om(k)t)}}
\def\ekdod{e^{i[\k\cdot(\x-\xp)-\om(t-t')]}}
\def\ekxxpa{e^{ik\xxpa}}
\def\qko{q(\k,\om)_}
\def\ekx{e^{i\k\cdot\x}}
\def\emkx{e^{-i\k\cdot\x}}
\def\ekz{e^{ikz}}
\def\emkz{e^{-ikz}}
\def\ekr{e^{ikr}}
\def\ekrp{e^{ikr'}}
\def\ekR{e^{ikR}}
\def\ekRp{e^{ikR'}}
\def\emknxp{e^{-ik(\nn\cdot\xp)}}
\def\emkxp{e^{-i\k\cdot\xp}}
\def\emkzot{e^{-i(kz+\om t)}}
\def\ekrot{e^{i(kr-\om t)}}
\def\uvk{u_\k}
\def\uk{u_k}
\def\lep{\left(}
\def\rip{\right)}
\def\Phx{\Ph(\x)}
\def\Phxt{\Ph(\x,t)}
\def\Phko{\Ph(\k,\om)}
\def\Phmx{\Ph_M(\x)}
\def\Phxp{\Ph(\xp)}
\def\phx{\ph(\x)}
\def\psx{\ps(\x)}
\def\fxt{f(\x,t)}
\def\Fxt{F(\x,t)}
\def\Fx{\F(\x)}
\def\Fxo{F(\x,\om)}
\def\Fko{F(\k,\om)}
\def\fxtp{f(\xp,t')}
\def\frvt{f(\r,\v,t)}
\def\Frt{F(\r,t)}
\def\psxt{\ps(\x,t)}
\def\phxp{\ph(\xp)}
\def\psxp{\ps(\xp)}
\def\Ak{A(k)}
\def\uzt{U(z,t)}
\def\Ex{\E(\x)}
\def\Ext{\E(\x,t)}
\def\Eo{\E(\om)}
\def\Exo{\E(\x,\om)}
\def\Eko{\E(\k,\om)}
\def\Dx{\D(\x)}
\def\Dxt{\D(\x,t)}
\def\Dxo{\D(\x,\om)}
\def\Px{\P(\x)}
\def\Pxt{\P(\x,t)}
\def\Pxo{\P(\x,\om)}
\def\Po{\P(\om)}
\def\Ax{\A(\x)}
\def\Axt{\A(\x,t)}
\def\Ako{\A(\k,\om)}
\def\Bx{\B(\x)}
\def\Bxt{\B(\x,t)}
\def\Bxo{\B(\x,\om)}
\def\Bko{\B(\k,\om)}
\def\Hx{\H(\x)}
\def\Hxt{\H(\x,t)}
\def\Mx{\M(\x)}
\def\Mxt{\M(\x,t)}
\def\Mxp{\M(\xp)}
\def\nrt{n(\r,t)}
\def\Jx{\J(\x)}
\def\Jxt{\J(\x,t)}
\def\Jko{\J(\k,\om)}
\def\Jxo{\J(\x,\om)}
\def\Jxop{\J(\x,\om')}
\def\Jxp{\J(\xp)}
\def\chx{\ch(\x)}
\def\chxt{\ch(\x,t)}
\def\v{{\bf v}}
\def\l{{\bf l}}
\def\b{{\bf b}}
\def\c{{\bf c}}
\def\u{{\bf u}}
\def\vfxt{{\bf f}(\x,t)}
\def\mm{{\bf \mu}}
\def\N{{\bf N}}
\def\eq#1{Eq.~(#1)}
\def\rhx{\rh(\x)}
\def\rhr{\rh(\vr)}
\def\rhxt{\rh(\x,t)}
\def\rhko{\rh(\k,\om)}
\def\rhxp{\rh(\xp)}
\def\rhxo{\rh(\x,\om)}
\def\rhxop{\rh(\x,\om')}
\def\leb{\left[}
\def\rib{\right]}
\def\six{\si(\x)}
\def\sixp{\si(\xp)}
\def\gxxp{G(\x,\xp)}
\def\gxt{G(\x,t;\xp,t')}
\def\gko{g(\k,\om)}
\def\dexxp{\de(\x-\xp)}
\def\dettp{\de(t-t')}
\def\dekkp{\de(\k-\k')}
\def\deoop{\de(\om-\om')}
\def\fxxp{F(\x,\xp)}
\def\lapp{\nabla'^2}
\def\xxpi{\frac{1}{|\x-\xp|}}
\def\xxpa{|\x-\xp|}
\def\xxp{(\x-\xp)}
\def\dvl{d{\bf l}}
\def\gradp{\nabla'}
\def\nnp{{\bf n}'}
\def\hsp#1{\hspace*{#1 in}}
\def\vsp#1{\vspace*{#1 in}}
\def\lel{\left|}
\def\ril{\right|}
\def\zh{\hat{{\bf z}}}
\def\xh{\hat{{\bf x}}}
\def\yh{\hat{{\bf y}}}
\def\p{{\bf p}}
\def\r{{\bf r}}
\def\eqs#1#2{Eqs.~(#1) and (#2)}
\def\eqss#1#2#3{Eqs.~(#1),~(#2), and~(#3)}
\def\eqsss#1#2#3#4{Eqs.~(#1), (#2), (#3), and~(#4)}
\def\sde#1#2{\frac{d^2#1}{d#2^2}}
\def\csde#1#2#3{\frac{d^2#1}{d#2 d#3}}
\def\y{{\bf y}}
\def\unet{U_n(\et)}
\def\umet{U_m(\et)}
\def\unets{U_n^*(\et)}
\def\umets{U_m^*(\et)}
\def\inabet{\int_a^bd\et}
\def\Ps{\Psi}
\def\sth{\sin\th}
\def\cth{\cos\th}
\def\sph{\sin\ph}
\def\cph{\cos\ph}
\def\inoou{\int_{-1}^1du}
\def\lec{\left\{}
\def\ric{\right\}}
\def\ylm{Y_{l,m}(\th,\ph)}
\def\ylms{Y_{l,m}^*(\th,\ph)}
\def\ylmsp{Y_{l,m}^*(\th',\ph')}
\def\ylmp{Y_{l'm',}(\th,\ph)}
\def\ylmps{Y_{l'm'}^*(\th,\ph)}
\def\inom{\int d\Om}
\def\les{\left/}
\def\led{\left.}
\def\rid{\right.}
\def\ris{\right/}
\def\Ga{\Gamma}
\def\De{\Delta}
\def\Th{\Theta}
\def\jnx{J_\nu(x)}
\def\nnx{N_\nu(x)}
\def\jnxy{J_\nu(x_{\nu n}y)}
\def\jxy{J_\nu(xy)}
\def\jxpy{J_\nu(x'y)}
\def\xnn{x_{\nu n}}
\def\jnxpy{J_\nu(x_{\nu n'}y}
\def\xnnp{x_{\nu n'}}
\def\nlb{\nolinebreak}
\def\edm{{\bf p}}
\def\mdm{{\bf m}}
\def\Pxp{{\bf P}(\xp)}
\def\rhh{\hat{{\bf r}}}
\def\xdo{\dot x}
\def\ydo{\dot y}
\def\zdo{\dot z}
\def\xdd{\ddot x}
\def\ydd{\ddot y}
\def\zdd{\ddot z}
\def\rhd{\dot\rh}
\def\rhdd{\ddot\rh}
\def\baselinestretch{1.7}
\title{\bf Molecular dynamics simulation of compressible fluid flow in
two-dimensional channels}
\author{M.~Sun and C.~Ebner\\Department of Physics \\
The Ohio State University\\Columbus, OH 43210}
\bdo
\maketitle
\bab
We study compressible fluid flow in narrow two-dimensional channels
using a novel molecular dynamics simulation method.
In the simulation area, an upstream source is maintained
at constant density and temperature while a downstream reservoir
is kept at vacuum. The channel is sufficiently long in the direction of the
flow that the finite length has little effect on the properties of the
fluid in the central region. The simulated system
is represented by an efficient data structure, whose internal
elements are created and manipulated dynamically in a layered
fashion. Consequently the code is highly efficient and manifests
completely linear performance in simulations of large systems.
We obtain the steady-state velocity, temperature, and density
distributions in the system. The velocity distribution across the channel
is very nearly a quadratic function of the distance from the center of the
channel and reveals velocity slip at the boundaries; the temperature
distribution is only approximately a quartic function of this distance
from the center to the channel. The density distribution across the channel is
non-uniform. We attribute this non-uniformity to the relatively high Mach
number, approximately 0.5, in the fluid flow. An equation for the
density distribution based on simple compressibility arguments is proposed;
its predictions agree well with the simulation results.
Validity of the concept of local dynamic temperature and the variation of
the temperature along the channel are discussed.

\eab
\section{Introduction}
The technique of molecular dynamics (MD) simulation has been widely used to
study non-equilibrium fluids. Because of limitations imposed by finite
computational capacity with regard to both memory and speed, this method has
been used principally to determine the behaviors of fluid systems on time and
distance scales within a few orders of magnitude of $ \tau_c$ and $r_c$,
respectively. Here, $\tau_c \approx 10^{-13}\,sec$ is the collision duration
and $r_c \approx 10^{-8}\,cm$ is a molecular size\cite{jpb}. Only molecular
properties of the fluid can be obtained in this range of time and distance, and
that is also the domain of experimental neutron scattering measurements.
Realistic examination of the hydrodynamic properties of flow is still beyond
the reach of most molecular dynamics simulations, although there are many
attempts to simulate larger systems on longer time scales. Among these are
the work of Koplik, \etal,$^{2,3}$ Hannon, \etal,\cite{lh174} and
Bhattacharya and Lie.\cite{dkb897,dkb761} In particular, Hannon, \etal, have
obtained velocity and temperature distributions across channels
in which flow is occurring. Their results agree well with simple
hydrodynamic predictions for incompressible fluid flow. Also, Bhattacharya and
Lie have obtained similar velocity profiles and have computed boundary slip
coefficients.

These simulations of fluid flow have a common property. Periodic boundary
conditions are introduced along the flow direction in the interest of
reducing the amount of computation needed to obtain useful results. The
imposition of translational invariance along the flow direction makes it
necessary to introduce a ``gravitational'' field to induce flow.
In order to induce appreciable flow, this field must be
given a strength much larger than the earth's field $g$, for example, as
large as\cite{jk781} $10^{12}g$. As a consequence of the gravitational field,
regular rescaling of the particles' kinetic energies is required; hence
one cannot reliably study properties of the fluid having to do with
energy or heat flow. Also, it is not possible to study variations of the flow
properties along the flow direction.

In this paper, we present a new method of simulation in which the
channel is of finite length in the flow direction without periodic boundary
conditions. At one end of the channel is a source region which is maintained at
constant density and temperature by introducing particles as needed.
At the other end is a sink region which is maintained at vacuum; that is, any
particle which moves into this region is removed from the system. Hence the
pressure or density gradient along the channel is primarily responsible for
instigation of the flow. We also use a novel layered data structure to
improve the efficiency of the code and the utilization of storage.
Our methods make it possible to simulate systems containing 20,000 or
more particles for more than $10^6$ time steps on a DECstation 3100.

Section II contains a description of the system simulated and of our
numerical techniques. The results are in Sec.~III, and Sec.~IV contains a
discussion and conclusions.
\section{Model and numerical methods}
The geometry of the system is indicated in Fig.~1. The channel has a length
$L$ in the $x$ direction and width $w$ in the $y$ direction. Flow is along
the $x$ direction with the source region at $0<x<L_1$ and the vacuum or
sink located at $x>L$. The channel is closed at $x=0$ so that particles
cannot escape into the region $x<0$. Typical channel sizes that we have
investigated are $L=400\si$, $w=100\si$, and $L_1=100\si$, where $\si$ is the
particle size parameter (diameter) in the Lennard-Jones potential
\beq
V(r)=4\ep\leb\lep\frac\si r\rip^{12}-\lep\frac\si r\rip^6\rib.
\eeq
which we have employed for the interparticle interactions. The interaction
is truncated at $r \ge 2\si$\cite{note}. In addition to
$\ep$ and $\si$, the only other parameter describing the properties of the
molecules is the mass $m$. We have used values appropriate for argon, \ie
$\si=3.4\,$\AA, $\ep/k=119.76\,K$, and $m=6.67\times10^{-23}\,g$; $k$ is the
Boltzmann constant. We also introduce a basic time constant $\tau=
\sqrt{m\si^2/48\ep}\approx3\times10^{-13}\,sec$.

Our procedure for handling collisions of the particles with the walls, which
are the surfaces $x=0$, $y=0$, and $y=w$, is to give the recoiling particles
a Maxwell-Boltzmann velocity distribution on a half-space. For example,
after colliding with the wall at $x=0$ a particle is given a velocity
$\v=v_x\xh+v_y\yh$ (where $v_x>0$) with a probability proportional to
$\exp[-m(v_x^2+v_y^2)/2kT]$ where $T$ is the specified wall temperature.
The molecular dynamics simulation proceeds in the conventional manner except
that when the particle density in the source region drops by a small amount,
typically after several hundred time steps, we inject particles into this
region in order to bring the density here back up to some prescribed value.
That is done by, first, letting the end wall ($x=0$) act as a piston and
uniformly compress the particles in the source region (but not
the remainder of the channel) by a sufficient amount to bring the density in
this compressed volume up to the desired level, and, second, restoring the
end wall to its original position and injecting an appropriate number of
particles into the empty space next to the wall. These added particles are
given a Maxwell velocity distribution at the same temperature $T$ as that of
the walls. By making this adjustment to the number of particles sufficiently
frequently, we only have to inject a few particles each time and the
relative amount by which the source region is compressed is very small. By
comparison, the entire system typically contains ten to twenty thousand
particles, and the source region is one-fourth of the channel.

The initial state of the system is always taken to have some preset density
of particles with a Maxwell-Boltzmann velocity distribution and no additional
particles elsewhere. Thus, as time proceeds, particles make their way
down the channel, driven by the pressure or density gradient, and are
removed at the far end. After a sufficiently long time a steady state is
achieved and various quantities such as the distributions of flow velocity,
particle density, particle current density, and temperature, may be computed.
We typically run the simulation in the steady state for some $10^6$ time steps
with one time step being $\de t=3\times10^{-2}\tau \approx9\times 10^{-15}\,
sec$. In addition, some $10^6$ time steps are needed for the system
to reach the steady state.

The most critical part of a molecular dynamics simulation is the
force calculation\cite{mpa} which would require $O(N^2)$ steps in a brute
force calculation if $N$ is the number of particles. Given short-ranged
interactions, such as the Lennard-Jones potential, one typically uses Verlet's
neighbor list\cite{mpa} to lower the number of steps to $O(NN_{c})$, where
$N_{c}$ is the maximum number of particles that can fit within the range of a
particle's potential. This method has high memory requirements and also
requires a number of steps $O(N^{2})$ to update the neighbor list. A
superior method\cite{mpa} in large systems is to divide the simulation area
into cells and use two arrays HEAD and LIST to store linked lists. This method
uses no extra memory and has linear performance, \ie $O(N)$ steps. However, the
reference locality is broken when molecules in a given cell are stored in a
long array LIST for a large system. This can be detrimental to the
performance of the code.

In our implementation of the MD simulation we
build a specially tailored data structure to reflect the similarity
between the physical problem and the memory partition of modern
computers. It is shown schematically in Fig.~2. A node, which
is roughly speaking a piece of computer memory, is created for every
particle introduced into the system. All the information for that particle is
stored in the node. The memory of the node is deallocated when the particle
moves out of the system. The system is divided into cells to reflect the
short-ranged nature of the interaction. The neighboring environment of every
molecule is implemented by a dynamic one-way link list in each cell.
This one-to-one dynamic management of the simulated systems enables a high
degree of reference locality and efficient usage of computer resources.

This data representation also enables layered structures.
We already see the atom/node layer and the cell layer. Further
layers can be built on top of these to take into account the structures of
\eg polyatomic molecules or other complex objects. Hence this method of
constructing the data improves the code reusability and makes simulation of
complex systems relatively easy. We have, for example, extended our code to
simulate the flow of a fluid composed of propane ($C_3H_8$) molecules.
With this data structure, we are able to achieve on a single DECstation
3100 a speed of $3\times10^{-5}$ second/particle/step, comparable or superior
to that obtained in some\cite{jk1282,pb789} but not all\cite{gre} simulations
done on supercomputers for comparable systems.
\section{Results}
All of the figures in this section are for steady-state flow with
$L=400\si$, $w=100\si=L_1$, and $T=\ep/k$ in the source region. The lone
exception is the inset in Fig.~5 which is for a narrow channel with $w=40\si$.
In all cases the fluid is maintained in the source region at a density
$n=0.25/\si^2$. The temperature is well above the critical temperature
$T_c\approx0.55\ep/k$ of the two-dimensional Lennard-Jones fluid\cite{ffa} and
the density is well below solid densities $n_s\ge0.8/\si^2$ so that we will
have a one-phase system unless sufficient cooling takes place at some point
along the channel to produce a much lower temperature. Some cooling is expected
as the fluid expands and moves down the channel; however, collisions with the
walls, which are also at $T=\ep/k$, will compensate somewhat for this effect.

Figure~3 shows the particle current density along the channel, $J_x$, in units
of $1/(\si\ta)$, as a function of $x/\si$. Different sets of points correspond
to different intervals of $y$. For example, the points labelled ``0-10''
represent an average of the current density for $0<y<10\si$, and of the
current density for $w-10\si<y<w$. Because the distribution is symmetric around
the line $y=w/2$, we can average over the data from strips symmetrically placed
on the two sides of this line. We have done so in order to obtain smoother
results. The current density averaged across the entire channel is also shown.
The constancy of this density outside of the source region, as a function of
$x$, is an indication that the system is indeed in the steady state. The
data in this figure, as well as in all of the following ones, are
obtained by averaging over a million time steps after discarding
results from an initial one million steps during which the steady state is
achieved. Notice also the linear increase of the current densities in the
source region.

In Fig.~4 we show the particles' mean velocity in the $x$ direction, $u$, in
units of $\si/\ta$, as a function of the position across the channel $y/w$.
The velocity is evaluated in region II of the channel which is the middle
third of the flow area excluding the source region. In this part of the
channel the end effects are minimized. The velocity $u$ is the macroscopic
fluid velocity. One may see clearly the velocity slip at the walls. Also
given in the figure as a solid line is a parabola which has been fit to the
data. For an incompressible fluid and laminar flow, the velocity profile is
expected to be parabolic. From the figure one can see that, even though we have
a highly compressible fluid, \cf,~Figs.~5 and~6, the distribution still fits
the parabola quite well.

The cross-channel density distribution in region II is shown in Fig.~5, and
the density along the channel, averaged across the channel, appears in
Fig.~6. One can see that $n$ decreases roughly, but definitely not
precisely, linearly as $x$ increases. Also, it is
is significantly lower in the middle of the channel than toward the
sides. The cross-channel variation is expected for compressible
fluid flow when the velocity is large enough.\cite{pat}
For Mach number less than about 0.3, the density variation is usually not great
but when it is around 0.5 or larger, which is the case in our system, the
finite compressibility of the fluid is highly evident. Exact solutions of the
hydrodynamic equations for the flow of a compressible fluid are known only for
a few special cases. We propose the following simple arguments to explain
qualitatively the density distribution observed in our simulations. First,
since the local temperature variation across the channel is quite small in
comparison with the total kinetic energy variation, we shall neglect the
former. Then to lowest order in the Mach number $M$ we have the following
generalized Bernoulli equation for isentropic flow,
\beq
P_0-P=\frac12\rh c^2M^2
\eeq
where $c$ is the speed of sound; $P$ is the pressure at a point where the flow
velocity is $u$; and $P_0$ is the pressure where $u=0$, or stagnation pressure.
The speed of sound can be approximated as follows,
\beq
c^2 =\pdes P\rh S\approx\lep\frac{\De P}{\De\rh}\rip_S,
\eeq
and we can identify $\De P=P-P_0$ and $\De\rh=\rh-\rh_0$ where $\rh_0$ is
the stagnation density. Then we have, still to second order in $M$,
\beq
\rh={\rh_0 \over {1+\frac12 M^2}}
={\rh_0 \over {1+\frac12 {u^2 \over c^2}}}
\eeq
This function is plotted as a solid line in Fig.~5 with $u$ computed from
the simulation. For $\rh_0$ we use the density at the downstream end of the
source region, $\th_0=0.23/\si^2$. We did not use the mean density in this
region because of the unphysical rescaling we use here in the simulations
to balance the extra heat generated by bringing in new particles.
In addition, the speed of sound is temperature-dependent and is not
a constant throughout the channel leading to a further ambiguity in
connection with the appropriate choice of $c$. If we simply choose $c$ to
be the ideal gas value $c=\sqrt{2kT/m}$ with $T=0.75/k$ which is the
average temperature in region II of the channel, \cf, fig.~8, that gives
the result shown in fig.~5. Qualitatively, the fit is quite reasonable,
perhaps better than one might have expected.

Also shown as an inset in Fig.~5 is the cross-channel density distribution
in region II for a considerably narrower channel with $w=40\si$. In this
case the density is quite constant across the channel except very close to
the edges where it decreases somewhat. By contrast, the cross-channel velocity
distribution for the narrow channel shows behavior very similar to that for
the side channel except that the peak velocity in the center is considerably
smaller. The flow is sufficiently slow that no density decrease is
apparent in the middle of the channel as compared with the density closer
to the sides at a given $x$.

For both the narrow channel and the wide one, there is a significant
decrease in the particle density very close to the walls, \ie, within
about $2\si$ of the walls. We believe this decrease to depend on, and be a
consequence of, the manner in which we reflect particles from a wall.

The local temperature distribution in the channel has been obtained from
study of the velocity distributions of the particles within each cell. Of
course, one may ask whether there is a well-defined temperature in an
intrinsically non-equilibrium system. We obtain the temperature from the
equipartition theorem according to which, in the absence of any flow,
$kT=m\ol{v_x^2}$ or $kT=m\ol{v_y^2}$ where the overlines indicated mean
values taken over the distribution of particle velocities; both temperatures
should be the same. However, given that there is some flow velocity $u$ in the
$x$ direction as in our channel, then if there is a Maxwell velocity
distribution centered at this velocity, we may introduce temperatures
\beq
kT_x=m\ol{(v_x-u)^2}\andh kT_y=m\ol{v_y^2}.
\eeq
Using these formulas we have computed $T_x$ and $T_y$. The cross-channel
temperature distributions in region II, in units of $\ep/k$, are shown
in Fig.~7. The two generally agree quite well except very close to the
edges of the channel where $T_x$ is considerably larger than $T_y$. One should
expect that they would not agree here because the velocity distribution
is certainly not a moving Maxwell distribution as a consequence of the
particular manner in which we handle reflections of the particles from the
walls. Rather, the distribution close to a wall is a superposition of two
distributions; the reflected particles have a Maxwell distribution which is not
moving while the incident particles have a net velocity along the $x$
direction. The solid curve in Fig.~7 is the best fit of a quartic function to
the data. A quartic has been used because that is the predicted temperature
distribution for an incompressible fluid. One can see that the fit is not
very good, indicating that the prediction of incompressible fluid theory is
really not appropriate in this case.

Figure~8 shows the average temperature distribution along the channel.
One can see that $T_x$ and $T_y$ agree quite well and that the temperature
drop, once the particles have left the source region, is roughly linear.
Several factors contribute to the temperature drop. First, the fluid is
expanding, which will increase its potential energy, and so to the extent
that energy is conserved the mean kinetic energy per particle must decrease.
Second, collisions with the walls offset this effect because particles
reflected from the walls have on the average a kinetic energy of $kT$ where
the wall temperature $T$ is the same as the initial temperature of particles
injected into the source region. Finally, there is also the effect of the
increasing flow velocity along the channel which has the consequence that,
as $x$ increases, more of the kinetic energy is going into the flow and
correspondingly less energy is available for fluctuations of the particles'
velocities relative to the flow velocity thereby producing a drop in the
temperature, as we have defined it, with increasing $x$.
\section{Conclusions}
We find that, in two dimensions at least, it is practical to simulate a
realistic flow system with a source and a drain, and at the same time keep a
long enough channel that finite size effects associated with the length
are not important in regions farthest from the source or drain. Sufficiently
high Mach numbers to allow for the investigation of compressible fluid flow can
be achieved as is evidenced by the considerable cross-channel density
variations observed. The non-uniformity of this density distribution can be
roughly explained by an equation based on the Bernoulli equation and the
relation between system compressibility and the sound velocity.
This equation also predicts, and our results verify, that the density
distribution will be quite flat for flow with small Mach numbers.

We also find that there is good agreement between our
results for the flow velocity and predictions based on the hydrodynamics of
incompressible fluids. Overall the results suggest that for fairly high ($M
\sim\frac12$), but subsonic, Mach number flow, a fairly good picture of the
system's behavior can be obtained from solutions for incompressible fluid flow
with some corrections of order $M^2$ to account for the consequences of
compressibility.\\
\ul{Acknowledgments}. We wish to thank Gary Grest for a useful discussion
concerning molecular dynamics simulations of fluids and for bringing
ref.~10 to our attention. This work was supported by National Science
Foundation Grant DMR-9014679.
\pagebreak
\bbi
\bibitem{jpb} J.~P.~Boon and S.~Yip, {\sl Molecular Hydrodynamics}
(McGraw-Hill Inc., New York, 1980).

\bibitem{jk781} J.~Koplik, J.~R.~BanxDavar and J.~F.~Willemsen, {\sl
Phys. Fluids}, A1, 781 (1989).

\bibitem{jk1282} J.~Koplik, J.~R.~BanxDavar and J.~F.~Willemsen, {\sl
Phys. Rev. Lett.}~{\bf 60}, 1282 (1988).

\bibitem{lh174} L.~Hannon, G.~C.~Lie, and E.~Clementi,
{\sl Phys.~Lett.}~{\bf A119}, 174 (1986).

\bibitem{dkb897} D.~K.~Bhattacharya and G.~C.~Lie,
{\sl Phys. Rev. Lett.}~{\bf 62}, 897 (1989).

\bibitem{dkb761} D.~K.~Bhattacharya and G.~C.~Lie, {\sl Phys.~Rev.~A}~{\bf
43}, 761 (1991).

\bibitem{note} ~For every particle in a cell, we only calculate its
interaction with particles in the same cell and neighboring ones.
Since each cell has a length of $2\si$, the cutoff distance of our
force calculation is at least $2\si$ in worst cases.

\bibitem{mpa} M.~P.~Allen and D.~J.~Tildesley, {\sl Computer Simulation
of Liquids} (Clarendon Press, Oxford, 1987).

\bibitem{pb789} P.~Borgelt, C.~Hoheisel and G.~Stell, {\sl Phys.
Rev.~A} {\bf 42}, 789 (1990).

\bibitem{gre} G.~S.~Grest, B.~D\"unweg, and K.~Kremer,
{\sl Comp.~Phys.~Commun.}~{\bf 55}, 269 (1989).

\bibitem{ffa} F.~F.~Abraham, {\sl Physics Reports} {\bf 80}, 340 (1981).

\bibitem{pat} P.~A.~Thompson, {\sl Compressible-fluid Dynamics}
(McGraw-Hill, Inc, New York, 1972).

\ebi
\newpage
\bce
{\bf Figure Captions}
\ece
\bit
\item[Fig.~1] Diagram of the two-dimensional channel showing the source and
sink regions. Walls are indicated by solid lines. Average cross-channel
distributions presented in some of the following figures are obtained in region
II which is equidistant from the two ends of the simulation area,
excluding the source region. The lengths of the channel in the $x$ and
$y$ directions are denoted in the text by $L$ and $w$, respectively; the length
of the source in the $x$ direction is called $L_1$.
\item[Fig.~2] Schematic representation of the data structures employed in
the molecular dynamics simulations. Each node stores all information
such as position, acceleration, etc., of a particle. The simulation
area is divided into rectangular cells. Particles (nodes) in each cell
are linked by pointers to form a dynamic link list.
\item[Fig.~3] The particle current density $J_x$, in units of $1/\si\ta$ and
averaged within various intervals $\De y/w$ across the channel as indicated
by the legends, is shown as a function of $x/\si$. Also shown as a solid line
is $J_x$ averaged across the entire channel. These results are for a channel
with $w=100\si$ and $L=400\si$; the source and wall temperature is $T=\ep/
k$; and the averages are taken over $10^6$ time steps after an initial
$10^6$ steps during which the system reaches the steady state.
\item[Fig.~4] The steady-state average fluid velocity $u$ in region II is
plotted against
$y/w$; $u$ is given in units of $\si/\ta$. The parameters are the same as
for Fig.~3. The solid line represents the best fit of a parabola to the
simulation results.
\item[Fig.~5] The steady-state cross-channel particle density $n$ in region II
in units of
$1/\si^2$ is shown as a function of $x/\si$ for a wide channel ($w=100\si$) as
a function of $y/w$. The solid line is the theory described in the text. The
inset displays the particle density for a narrow channel ($w=40\si$). Other
parameters are the same as for fig.~3.
\item[Fig.~6] The steady-state average density along the channel in units of
$1/\si^2$ is shown as a function of $x/\si$. Parameters are the same as in
Fig.~3.
\item[Fig.~7] The steady-state cross-channel temperature distributions $T_x$
and $T_y$ in region II, extracted from the equipartition theorem by computing
the particles' kinetic energies, are given in units of $\ep/k$ as functions of
$y/w$. Parameters are the same as for Fig.~3.
\item[Fig.~8] The steady-state temperatures $T_x$ and $T_y$ in units of
$\ep/k$, averaged across the channel, are shown as functions of $x/\si$.
Parameters are the same as for Fig.~3.
\eit
\edo


/TeXDict 200 dict def TeXDict begin /N /def load def /B{bind def}N /S /exch
load def /X{S N}B /TR /translate load N /isls false N /vsize 10 N /@rigin{
isls{[0 1 -1 0 0 0]concat}if 72 Resolution div 72 VResolution div neg scale
Resolution VResolution vsize neg mul TR}B /@letter{/vsize 10 N}B /@landscape{
/isls true N /vsize -1 N}B /@a4{/vsize 10.6929133858 N}B /@a3{/vsize 15.5531 N
}B /@ledger{/vsize 16 N}B /@legal{/vsize 13 N}B /@manualfeed{statusdict
/manualfeed true put}B /@copies{/#copies X}B /FMat[1 0 0 -1 0 0]N /FBB[0 0 0 0
]N /df{/sf 1 N /fntrx FMat N df-tail}B /dfs{div /sf X /fntrx[sf 0 0 sf neg 0 0
]N df-tail}B /df-tail{/nn 8 dict N nn begin /FontType 3 N /FontMatrix fntrx N
/FontBBox FBB N string /base X array /BitMaps X /BuildChar{CharBuilder}N
/Encoding IE N end dup{/foo setfont}2 array copy cvx N load 0 nn put /ctr 0 N[
}B /E{pop nn dup definefont setfont}B /ch-image{ch-data dup type /stringtype
ne{ctr get /ctr ctr 1 add N}if}B /ch-width{ch-data dup length 5 sub get}B
/ch-height{ch-data dup length 4 sub get}B /ch-xoff{128 ch-data dup length 3
sub get sub}B /ch-yoff{ch-data dup length 2 sub get 127 sub}B /ch-dx{ch-data
dup length 1 sub get}B /ctr 0 N /CharBuilder{save 3 1 roll S dup /base get 2
index get S /BitMaps get S get /ch-data X pop /ctr 0 N ch-dx 0 ch-xoff ch-yoff
ch-height sub ch-xoff ch-width add ch-yoff setcachedevice ch-width ch-height
true[1 0 0 -1 -.1 ch-xoff sub ch-yoff .1 add]{ch-image}imagemask restore}B /D{
/cc X dup type /stringtype ne{]}if nn /base get cc ctr put nn /BitMaps get S
ctr S sf 1 ne{dup dup length 1 sub dup 2 index S get sf div put}if put /ctr
ctr 1 add N}B /I{cc 1 add D}B /bop{userdict /bop-hook known{bop-hook}if /SI
save N @rigin 0 0 moveto}B /eop{clear SI restore showpage userdict /eop-hook
known{eop-hook}if}B /@start{userdict /start-hook known{start-hook}if
/VResolution X /Resolution X 1000 div /DVImag X /IE 256 array N 0 1 255{IE S 1
string dup 0 3 index put cvn put}for}B /p /show load N /RMat[1 0 0 -1 0 0]N
/BDot 8 string N /v{/ruley X /rulex X V}B /V{gsave TR -.1 -.1 TR rulex ruley
scale 1 1 false RMat{BDot}imagemask grestore}B /a{moveto}B /delta 0 N /tail{
dup /delta X 0 rmoveto}B /M{S p delta add tail}B /b{S p tail}B /c{-4 M}B /d{
-3 M}B /e{-2 M}B /f{-1 M}B /g{0 M}B /h{1 M}B /i{2 M}B /j{3 M}B /k{4 M}B /l{p
-4 w}B /m{p -3 w}B /n{p -2 w}B /o{p -1 w}B /q{p 1 w}B /r{p 2 w}B /s{p 3 w}B /t
{p 4 w}B /w{0 rmoveto}B /x{0 S rmoveto}B /y{3 2 roll p a}B /bos{/SS save N}B
/eos{clear SS restore}B end
TeXDict begin /SDict 200 dict N SDict begin /@SpecialDefaults{/hs 612 N /vs
792 N /ho 0 N /vo 0 N /hsc 1 N /vsc 1 N /ang 0 N /CLIP false N /BBcalc false N
/p 3 def}B /@scaleunit 100 N /@hscale{@scaleunit div /hsc X}B /@vscale{
@scaleunit div /vsc X}B /@hsize{/hs X /CLIP true N}B /@vsize{/vs X /CLIP true
N}B /@hoffset{/ho X}B /@voffset{/vo X}B /@angle{/ang X}B /@rwi{10 div /rwi X}
B /@llx{/llx X}B /@lly{/lly X}B /@urx{/urx X}B /@ury{/ury X /BBcalc true N}B
/magscale true def end /@MacSetUp{userdict /md known{userdict /md get type
/dicttype eq{md begin /letter{}N /note{}N /legal{}N /od{txpose 1 0 mtx
defaultmatrix dtransform S atan/pa X newpath clippath mark{transform{
itransform moveto}}{transform{itransform lineto}}{6 -2 roll transform 6 -2
roll transform 6 -2 roll transform{itransform 6 2 roll itransform 6 2 roll
itransform 6 2 roll curveto}}{{closepath}}pathforall newpath counttomark array
astore /gc xdf pop ct 39 0 put 10 fz 0 fs 2 F/|______Courier fnt invertflag{
PaintBlack}if}N /txpose{pxs pys scale ppr aload pop por{noflips{pop S neg S TR
pop 1 -1 scale}if xflip yflip and{pop S neg S TR 180 rotate 1 -1 scale ppr 3
get ppr 1 get neg sub neg ppr 2 get ppr 0 get neg sub neg TR}if xflip yflip
not and{pop S neg S TR pop 180 rotate ppr 3 get ppr 1 get neg sub neg 0 TR}if
yflip xflip not and{ppr 1 get neg ppr 0 get neg TR}if}{noflips{TR pop pop 270
rotate 1 -1 scale}if xflip yflip and{TR pop pop 90 rotate 1 -1 scale ppr 3 get
ppr 1 get neg sub neg ppr 2 get ppr 0 get neg sub neg TR}if xflip yflip not
and{TR pop pop 90 rotate ppr 3 get ppr 1 get neg sub neg 0 TR}if yflip xflip
not and{TR pop pop 270 rotate ppr 2 get ppr 0 get neg sub neg 0 S TR}if}
ifelse scaleby96{ppr aload pop 4 -1 roll add 2 div 3 1 roll add 2 div 2 copy
TR .96 dup scale neg S neg S TR}if}N /cp{pop pop showpage pm restore}N end}if}
if}N /normalscale{Resolution 72 div VResolution 72 div neg scale magscale{
DVImag dup scale}if}N /psfts{S 65536 div N}N /startTexFig{/psf$SavedState save
N userdict maxlength dict begin /magscale false def normalscale currentpoint
TR /psf$ury psfts /psf$urx psfts /psf$lly psfts /psf$llx psfts /psf$y psfts
/psf$x psfts currentpoint /psf$cy X /psf$cx X /psf$sx psf$x psf$urx psf$llx
sub div N /psf$sy psf$y psf$ury psf$lly sub div N psf$sx psf$sy scale psf$cx
psf$sx div psf$llx sub psf$cy psf$sy div psf$ury sub TR /showpage{}N
/erasepage{}N /copypage{}N @MacSetUp}N /doclip{psf$llx psf$lly psf$urx psf$ury
currentpoint 6 2 roll newpath 4 copy 4 2 roll moveto 6 -1 roll S lineto S
lineto S lineto closepath clip newpath moveto}N /endTexFig{end psf$SavedState
restore}N /
{grestore
clear SpecialSave restore end}B /@defspecial{SDict begin}B /@fedspecial{end}B
/li{lineto}B /rl{rlineto}B /rc{rcurveto}B /np{/SaveX currentpoint /SaveY X N 1
setlinecap newpath}B /st{stroke SaveX SaveY moveto}B /fil{fill SaveX SaveY
moveto}B /ellipse{/endangle X /startangle X /yrad X /xrad X /savematrix matrix
currentmatrix N TR xrad yrad scale 0 0 1 startangle endangle arc savematrix
setmatrix}B end
TeXDict begin 1728 300 300 @start /Fa 12 106 df<3C7EFFFFFFFF7E3C08087A8714>46
D<000C00003C00007C0001FC003FFC00FE7C00C07C00007C00007C00007C00007C00007C00007C
00007C00007C00007C00007C00007C00007C00007C00007C00007C00007C00007C00007C00007C
00007C00007C00007C00007C00007C00007C00007C00007C00007C00007C00007C00007C00007C
00007C00007C00007C00007C00007C0000FE00FFFFFEFFFFFE172F7AAE24>49
D<007F800001FFE0000781F8000C007C0018003E0030003F0030001F8060000FC078000FC0FE00
0FC0FF000FE0FF000FE0FF0007E0FF0007E07E0007E03C000FE000000FE000000FC000000FC000
000FC000001F8000001F0000003F0000003E0000007C000000F8000000F0000001E0000003C000
0007800000070000000E0000001C0000003800000070000000E0006000C0006001800060030000
C0060000C00C0000C0180001C01FFFFFC03FFFFFC07FFFFF80FFFFFF80FFFFFF801B2F7CAE24>
I<003FC00000FFF80003C07E0007001F000C001F8018000FC01F000FC03F800FE03F8007E03F80
07E03F8007E01F800FE00F000FE000000FC000000FC000000F8000001F8000001F0000003E0000
007C000001F000007FC000007FC0000000780000001E0000000F0000000F80000007C0000007E0
000007F0000003F0000003F8000003F8180003F87E0003F8FF0003F8FF0003F8FF0003F8FF0003
F0FE0007F07C0007E0600007E030000FC018000F800E001F0007C07E0001FFF800003FC0001D30
7DAE24>I<000007000000070000000F0000001F0000001F0000003F0000007F0000007F000000
DF000001DF0000019F0000031F0000071F00000E1F00000C1F00001C1F0000381F0000301F0000
701F0000E01F0000C01F0001801F0003801F0003001F0006001F000E001F000C001F0018001F00
38001F0030001F0060001F00E0001F00FFFFFFFEFFFFFFFE00001F0000001F0000001F0000001F
0000001F0000001F0000001F0000001F0000001F0000001F0000003F80000FFFFE000FFFFE1F2F
7EAE24>I<180001801F801F801FFFFF001FFFFE001FFFFC001FFFF0001FFFC000187C00001800
000018000000180000001800000018000000180000001800000018000000180000001800000018
3F800018FFF00019C0F8001B003C001E001E001C000F0008000F8000000F80000007C0000007C0
000007C0000007E0000007E0000007E03C0007E07E0007E0FE0007E0FE0007E0FE0007E0FE0007
C0FC0007C060000F8060000F8030000F0030001E0018003E000E0078000781F00003FFC000007F
00001B307CAE24>I<0001FC00000FFF00003E0380007801C000F001C001E007E003C00FE00780
0FE00F800FE00F000FE01F0007C01E0000003E0000003E0000003E0000007C0000007C0100007C
0FF000FC3FFC00FC601E00FCC00F00FC800780FD0003C0FF0003E0FE0003E0FE0001F0FE0001F0
FC0001F0FC0001F8FC0001F8FC0001F8FC0001F87C0001F87C0001F87C0001F87C0001F83C0001
F83E0001F03E0001F01E0003E01E0003E00F0003C00F0007800780070003C00E0001F03C00007F
F800001FC0001D307DAE24>I<300000003C0000003FFFFFFC3FFFFFFC3FFFFFF87FFFFFF07FFF
FFF070000060600000C06000018060000180C0000300C0000600C000060000000C000000180000
00300000003000000060000000E0000000C0000001C00000018000000380000007800000070000
00070000000F0000000F0000001F0000001F0000003E0000003E0000003E0000003E0000007E00
00007E0000007E0000007E000000FE000000FE000000FE000000FE000000FE000000FE000000FE
000000FE0000007C0000003800001E317CAF24>I<001FC00000FFF80001E07C0003801F000700
07800E0007801E0003C01C0003C03C0001E03C0001E03C0001E03C0001E03E0001E03E0001C03F
0003C01FC003801FE007800FF80F0007FC1E0003FF380001FFF00000FFE000003FF000007FFC00
01E7FE0003C3FF800700FFC00E007FC01E001FE03C0007F0780003F0780001F8700000F8F00000
F8F0000078F0000078F0000078F0000078F000007078000070780000F03C0000E03C0001C01E00
03800F800F0003E03E0000FFF800003FC0001D307DAE24>I<FFFFFFFFFCFFFFFFFFFC03FC0003
FC01F800007C01F800003E01F800001E01F800000E01F800000E01F800000601F800000601F800
000601F800000601F800000301F800000301F800000301F800300301F800300001F800300001F8
00300001F800300001F800700001F800700001F801F00001FFFFF00001FFFFF00001F801F00001
F800700001F800700001F800300001F800300001F800300001F800300001F800300001F8000000
01F800000001F800000001F800000001F800000001F800000001F800000001F800000001F80000
0001F800000001F800000001F800000001F800000003FE000000FFFFFC0000FFFFFC000028317D
B02F>70 D<0000003C003F80FE00FFE38F03E0FE0F07803C0F0F001E0E1F001F001E000F001E00
0F003E000F803E000F803E000F803E000F803E000F801E000F001E000F001F001F000F001E0007
803C0007E0F8000CFFE000083F800008000000180000001C0000000C0000000E0000000FFFFC00
07FFFF8003FFFFE007FFFFF01E0003F83C0000FC7800003C7800003EF000001EF000001EF00000
1EF000001EF000001E7800003C3C0000781E0000F00F0001E007E00FC001FFFF00003FF800202F
7E9F24>103 D<038007C00FE01FE01FE00FE007C0038000000000000000000000000000000000
0000000003E07FE07FE007E003E003E003E003E003E003E003E003E003E003E003E003E003E003
E003E003E003E003E003E003E003E003E003E003E007F0FFFFFFFF10317FB014>105
D E end
TeXDict begin @letter /letter where {pop letter} if
bop 975 179 a 
 143 2656 a Fa(Fig.)32 b(8)p eop
bop 975 179 a 
 143 2656 a Fa(Fig.)32 b(7)p eop
bop 975 179 a 
 143 2656 a Fa(Fig.)32 b(6)p eop
bop 975 179 a 
 143 2656 a Fa(Fig.)32 b(5)p eop
bop 975 179 a 
 143 2656 a Fa(Fig.)32 b(4)p eop
bop 975 179 a 
 143 2656 a Fa(Fig.)32 b(3)p eop
bop 300 79 a
 21313290 22805214 2697052 13814169 29470228 42626580 startTexFig
300 79 a

/IdrawDict 53 dict def
IdrawDict begin

/reencodeISO {
dup dup findfont dup length dict begin
{ 1 index /FID ne { def }{ pop pop } ifelse } forall
/Encoding ISOLatin1Encoding def
currentdict end definefont
} def

/ISOLatin1Encoding [
/.notdef/.notdef/.notdef/.notdef/.notdef/.notdef/.notdef/.notdef
/.notdef/.notdef/.notdef/.notdef/.notdef/.notdef/.notdef/.notdef
/.notdef/.notdef/.notdef/.notdef/.notdef/.notdef/.notdef/.notdef
/.notdef/.notdef/.notdef/.notdef/.notdef/.notdef/.notdef/.notdef
/space/exclam/quotedbl/numbersign/dollar/percent/ampersand/quoteright
/parenleft/parenright/asterisk/plus/comma/minus/period/slash
/zero/one/two/three/four/five/six/seven/eight/nine/colon/semicolon
/less/equal/greater/question/at/A/B/C/D/E/F/G/H/I/J/K/L/M/N
/O/P/Q/R/S/T/U/V/W/X/Y/Z/bracketleft/backslash/bracketright
/asciicircum/underscore/quoteleft/a/b/c/d/e/f/g/h/i/j/k/l/m
/n/o/p/q/r/s/t/u/v/w/x/y/z/braceleft/bar/braceright/asciitilde
/.notdef/.notdef/.notdef/.notdef/.notdef/.notdef/.notdef/.notdef
/.notdef/.notdef/.notdef/.notdef/.notdef/.notdef/.notdef/.notdef
/.notdef/dotlessi/grave/acute/circumflex/tilde/macron/breve
/dotaccent/dieresis/.notdef/ring/cedilla/.notdef/hungarumlaut
/ogonek/caron/space/exclamdown/cent/sterling/currency/yen/brokenbar
/section/dieresis/copyright/ordfeminine/guillemotleft/logicalnot
/hyphen/registered/macron/degree/plusminus/twosuperior/threesuperior
/acute/mu/paragraph/periodcentered/cedilla/onesuperior/ordmasculine
/guillemotright/onequarter/onehalf/threequarters/questiondown
/Agrave/Aacute/Acircumflex/Atilde/Adieresis/Aring/AE/Ccedilla
/Egrave/Eacute/Ecircumflex/Edieresis/Igrave/Iacute/Icircumflex
/Idieresis/Eth/Ntilde/Ograve/Oacute/Ocircumflex/Otilde/Odieresis
/multiply/Oslash/Ugrave/Uacute/Ucircumflex/Udieresis/Yacute
/Thorn/germandbls/agrave/aacute/acircumflex/atilde/adieresis
/aring/ae/ccedilla/egrave/eacute/ecircumflex/edieresis/igrave
/iacute/icircumflex/idieresis/eth/ntilde/ograve/oacute/ocircumflex
/otilde/odieresis/divide/oslash/ugrave/uacute/ucircumflex/udieresis
/yacute/thorn/ydieresis
] def
/Helvetica-Bold reencodeISO def
/Helvetica reencodeISO def
/Times-Bold reencodeISO def

/arrowHeight 8 def
/arrowWidth 4 def
/none null def
/numGraphicParameters 17 def
/stringLimit 65535 def

/Begin {
save
numGraphicParameters dict begin
} def

/End {
end
restore
} def

/SetB {
dup type /nulltype eq {
pop
false /brushRightArrow idef
false /brushLeftArrow idef
true /brushNone idef
} {
/brushDashOffset idef
/brushDashArray idef
0 ne /brushRightArrow idef
0 ne /brushLeftArrow idef
/brushWidth idef
false /brushNone idef
} ifelse
} def

/SetCFg {
/fgblue idef
/fggreen idef
/fgred idef
} def

/SetCBg {
/bgblue idef
/bggreen idef
/bgred idef
} def

/SetF {
/printSize idef
/printFont idef
} def

/SetP {
dup type /nulltype eq {
pop true /patternNone idef
} {
dup -1 eq {
/patternGrayLevel idef
/patternString idef
} {
/patternGrayLevel idef
} ifelse
false /patternNone idef
} ifelse
} def

/BSpl {
0 begin
storexyn
newpath
n 1 gt {
0 0 0 0 0 0 1 1 true subspline
n 2 gt {
0 0 0 0 1 1 2 2 false subspline
1 1 n 3 sub {
/i exch def
i 1 sub dup i dup i 1 add dup i 2 add dup false subspline
} for
n 3 sub dup n 2 sub dup n 1 sub dup 2 copy false subspline
} if
n 2 sub dup n 1 sub dup 2 copy 2 copy false subspline
patternNone not brushLeftArrow not brushRightArrow not and and { ifill } if
brushNone not { istroke } if
0 0 1 1 leftarrow
n 2 sub dup n 1 sub dup rightarrow
} if
end
} dup 0 4 dict put def

/Circ {
newpath
0 360 arc
patternNone not { ifill } if
brushNone not { istroke } if
} def

/CBSpl {
0 begin
dup 2 gt {
storexyn
newpath
n 1 sub dup 0 0 1 1 2 2 true subspline
1 1 n 3 sub {
/i exch def
i 1 sub dup i dup i 1 add dup i 2 add dup false subspline
} for
n 3 sub dup n 2 sub dup n 1 sub dup 0 0 false subspline
n 2 sub dup n 1 sub dup 0 0 1 1 false subspline
patternNone not { ifill } if
brushNone not { istroke } if
} {
Poly
} ifelse
end
} dup 0 4 dict put def

/Elli {
0 begin
newpath
4 2 roll
translate
scale
0 0 1 0 360 arc
patternNone not { ifill } if
brushNone not { istroke } if
end
} dup 0 1 dict put def

/Line {
0 begin
2 storexyn
newpath
x 0 get y 0 get moveto
x 1 get y 1 get lineto
brushNone not { istroke } if
0 0 1 1 leftarrow
0 0 1 1 rightarrow
end
} dup 0 4 dict put def

/MLine {
0 begin
storexyn
newpath
n 1 gt {
x 0 get y 0 get moveto
1 1 n 1 sub {
/i exch def
x i get y i get lineto
} for
patternNone not brushLeftArrow not brushRightArrow not and and { ifill } if
brushNone not { istroke } if
0 0 1 1 leftarrow
n 2 sub dup n 1 sub dup rightarrow
} if
end
} dup 0 4 dict put def

/Poly {
3 1 roll
newpath
moveto
-1 add
{ lineto } repeat
closepath
patternNone not { ifill } if
brushNone not { istroke } if
} def

/Rect {
0 begin
/t exch def
/r exch def
/b exch def
/l exch def
newpath
l b moveto
l t lineto
r t lineto
r b lineto
closepath
patternNone not { ifill } if
brushNone not { istroke } if
end
} dup 0 4 dict put def

/Text {
ishow
} def

/idef {
dup where { pop pop pop } { exch def } ifelse
} def

/ifill {
0 begin
gsave
patternGrayLevel -1 ne {
fgred bgred fgred sub patternGrayLevel mul add
fggreen bggreen fggreen sub patternGrayLevel mul add
fgblue bgblue fgblue sub patternGrayLevel mul add setrgbcolor
eofill
} {
eoclip
originalCTM setmatrix
pathbbox /t exch def /r exch def /b exch def /l exch def
/w r l sub ceiling cvi def
/h t b sub ceiling cvi def
/imageByteWidth w 8 div ceiling cvi def
/imageHeight h def
bgred bggreen bgblue setrgbcolor
eofill
fgred fggreen fgblue setrgbcolor
w 0 gt h 0 gt and {
l b translate w h scale
w h true [w 0 0 h neg 0 h] { patternproc } imagemask
} if
} ifelse
grestore
end
} dup 0 8 dict put def

/istroke {
gsave
brushDashOffset -1 eq {
[] 0 setdash
1 setgray
} {
brushDashArray brushDashOffset setdash
fgred fggreen fgblue setrgbcolor
} ifelse
brushWidth setlinewidth
originalCTM setmatrix
stroke
grestore
} def

/ishow {
0 begin
gsave
fgred fggreen fgblue setrgbcolor
printFont printSize scalefont setfont
/descender 0 printFont /FontBBox get 1 get printFont /FontMatrix
get transform exch pop def
/vertoffset 1 printSize sub descender sub def {
0 vertoffset moveto show
/vertoffset vertoffset printSize sub def
} forall
grestore
end
} dup 0 2 dict put def

/patternproc {
0 begin
/patternByteLength patternString length def
/patternHeight patternByteLength 8 mul sqrt cvi def
/patternWidth patternHeight def
/patternByteWidth patternWidth 8 idiv def
/imageByteMaxLength imageByteWidth imageHeight mul
stringLimit patternByteWidth sub min def
/imageMaxHeight imageByteMaxLength imageByteWidth idiv patternHeight idiv
patternHeight mul patternHeight max def
/imageHeight imageHeight imageMaxHeight sub store
/imageString imageByteWidth imageMaxHeight mul patternByteWidth add string def
0 1 imageMaxHeight 1 sub {
/y exch def
/patternRow y patternByteWidth mul patternByteLength mod def
/patternRowString patternString patternRow patternByteWidth getinterval def
/imageRow y imageByteWidth mul def
0 patternByteWidth imageByteWidth 1 sub {
/x exch def
imageString imageRow x add patternRowString putinterval
} for
} for
imageString
end
} dup 0 12 dict put def

/min {
dup 3 2 roll dup 4 3 roll lt { exch } if pop
} def

/max {
dup 3 2 roll dup 4 3 roll gt { exch } if pop
} def

/arrowhead {
0 begin
transform originalCTM itransform
/taily exch def
/tailx exch def
transform originalCTM itransform
/tipy exch def
/tipx exch def
/dy tipy taily sub def
/dx tipx tailx sub def
/angle dx 0 ne dy 0 ne or { dy dx atan } { 90 } ifelse def
gsave
originalCTM setmatrix
tipx tipy translate
angle rotate
newpath
0 0 moveto
arrowHeight neg arrowWidth 2 div lineto
arrowHeight neg arrowWidth 2 div neg lineto
closepath
patternNone not {
originalCTM setmatrix
/padtip arrowHeight 2 exp 0.25 arrowWidth 2 exp mul add sqrt brushWidth mul
arrowWidth div def
/padtail brushWidth 2 div def
tipx tipy translate
angle rotate
padtip 0 translate
arrowHeight padtip add padtail add arrowHeight div dup scale
arrowheadpath
ifill
} if
brushNone not {
originalCTM setmatrix
tipx tipy translate
angle rotate
arrowheadpath
istroke
} if
grestore
end
} dup 0 9 dict put def

/arrowheadpath {
newpath
0 0 moveto
arrowHeight neg arrowWidth 2 div lineto
arrowHeight neg arrowWidth 2 div neg lineto
closepath
} def

/leftarrow {
0 begin
y exch get /taily exch def
x exch get /tailx exch def
y exch get /tipy exch def
x exch get /tipx exch def
brushLeftArrow { tipx tipy tailx taily arrowhead } if
end
} dup 0 4 dict put def

/rightarrow {
0 begin
y exch get /tipy exch def
x exch get /tipx exch def
y exch get /taily exch def
x exch get /tailx exch def
brushRightArrow { tipx tipy tailx taily arrowhead } if
end
} dup 0 4 dict put def

/midpoint {
0 begin
/y1 exch def
/x1 exch def
/y0 exch def
/x0 exch def
x0 x1 add 2 div
y0 y1 add 2 div
end
} dup 0 4 dict put def

/thirdpoint {
0 begin
/y1 exch def
/x1 exch def
/y0 exch def
/x0 exch def
x0 2 mul x1 add 3 div
y0 2 mul y1 add 3 div
end
} dup 0 4 dict put def

/subspline {
0 begin
/movetoNeeded exch def
y exch get /y3 exch def
x exch get /x3 exch def
y exch get /y2 exch def
x exch get /x2 exch def
y exch get /y1 exch def
x exch get /x1 exch def
y exch get /y0 exch def
x exch get /x0 exch def
x1 y1 x2 y2 thirdpoint
/p1y exch def
/p1x exch def
x2 y2 x1 y1 thirdpoint
/p2y exch def
/p2x exch def
x1 y1 x0 y0 thirdpoint
p1x p1y midpoint
/p0y exch def
/p0x exch def
x2 y2 x3 y3 thirdpoint
p2x p2y midpoint
/p3y exch def
/p3x exch def
movetoNeeded { p0x p0y moveto } if
p1x p1y p2x p2y p3x p3y curveto
end
} dup 0 17 dict put def

/storexyn {
/n exch def
/y n array def
/x n array def
n 1 sub -1 0 {
/i exch def
y i 3 2 roll put
x i 3 2 roll put
} for
} def

Begin
[ 0.923077 0 0 0.923077 0 0 ] concat
/originalCTM matrix currentmatrix def

Begin 
1 0 0 [2 2 2 2 2 2 2 2] 15 SetB
0 0 0 SetCFg
1 1 1 SetCBg
1 SetP
[ 1 0 0 1 32 136 ] concat
16 146 451 396 Rect
End

Begin 
1 0 0 [] 0 SetB
0 0 0 SetCFg
1 1 1 SetCBg
1 SetP
[ 1 0 0 1 265 -106 ] concat
75 490 167 533 Rect
End

Begin 
1 0 0 [] 0 SetB
0 0 0 SetCFg
1 1 1 SetCBg
1 SetP
[ 1 0 0 1 286 -32 ] concat
75 490 167 533 Rect
End

Begin 
[ 0 1 -1 0 680 372 ] concat

Begin 
Helvetica-Bold 14 SetF
[ 0 -1 1 0 -145 700 ] concat

Begin 
1 0 0 [] 0 SetB
0 0 0 SetCFg
1 1 1 SetCBg
1 SetP
[ 1 0 0 1 98 130 ] concat
74 243 285 342 Rect
End

Begin 
1 0 0 [] 0 SetB
0 0 0 SetCFg
1 1 1 SetCBg
1 SetP
[ 1 0 0 1 42 128.5 ] concat
169 342 169 245 Line
End

Begin 
1 0 0 [] 0 SetB
0 0 0 SetCFg
1 1 1 SetCBg
1 SetP
[ 1 0 0 1 159.5 129 ] concat
169 342 169 245 Line
End

Begin 
1 0 0 [] 0 SetB
0 0 0 SetCFg
1 1 1 SetCBg
1 SetP
[ 1 0 0 1 81 128.5 ] concat
169 342 169 245 Line
End

Begin 
1 0 0 [] 0 SetB
0 0 0 SetCFg
1 1 1 SetCBg
1 SetP
[ 1 0 0 1 119 129 ] concat
169 342 169 245 Line
End

End 

Begin 
0 0 0 SetCFg
Helvetica 12 SetF
[ 1 0 0 1 261.5 470 ] concat
[
(old)
] Text
End

Begin 
0 0 0 SetCFg
Helvetica-Bold 14 SetF
[ 1 0 0 1 250 437 ] concat
[
(a)
] Text
End

Begin 
0 0 0 SetCFg
Helvetica 14 SetF
[ 1 0 0 1 251.5 395.5 ] concat
[
(index)
] Text
End

Begin 
0 0 0 SetCFg
Helvetica 14 SetF
[ 1 0 0 1 251 353.5 ] concat
[
(pointer)
] Text
End

Begin 
0 0 0 SetCFg
Helvetica-Bold 14 SetF
[ 1 0 0 1 252 511.5 ] concat
[
(x)
] Text
End

Begin 
0 0 0 SetCFg
Helvetica-Bold 14 SetF
[ 1 0 0 1 251 476 ] concat
[
(x)
] Text
End

End 

Begin 
0 0 0 SetCFg
Times-Bold 14 SetF
[ 1 0 0 1 189 574 ] concat
[
(Individual Molecule (Node))
] Text
End

Begin 
0 0 0 SetCFg
Times-Bold 14 SetF
[ 1 0 0 1 177 244 ] concat
[
(Molecule Chain in A Cell)
] Text
End

Begin 
0 0 0 SetCFg
Times-Bold 14 SetF
[ 1 0 0 1 383 486 ] concat
[
(2)
] Text
End

Begin 
0 0 0 SetCFg
Times-Bold 14 SetF
[ 1 0 0 1 363 410 ] concat
[
(3)
] Text
End

Begin 
[ 1 0 0 1 177 -41 ] concat

Begin 
1 0 0 [] 0 SetB
0 0 0 SetCFg
1 1 1 SetCBg
1 SetP
[ 1 0 0 1 114 -137 ] concat
75 490 167 533 Rect
End

Begin 
0 0 0 SetCFg
Times-Bold 14 SetF
[ 1 0 0 1 210 382 ] concat
[
(4)
] Text
End

End 

Begin 
[ 1 0 0 1 -27 -61 ] concat

Begin 
1 0 0 [] 0 SetB
0 0 0 SetCFg
1 1 1 SetCBg
1 SetP
[ 1 0 0 1 149 -2 ] concat
75 490 167 533 Rect
End

Begin 
1 0 0 [] 0 SetB
0 0 0 SetCFg
1 1 1 SetCBg
1 SetP
[ 1 0 0 1 26 -31 ] concat
75 490 167 533 Rect
End

Begin 
0 0 0 SetCFg
Times-Bold 14 SetF
[ 1 0 0 1 119 485 ] concat
[
(Header)
] Text
End

Begin 
0 0 0 SetCFg
Times-Bold 14 SetF
[ 1 0 0 1 247 512 ] concat
[
(1)
] Text
End

Begin 
1 0 1 [] 0 SetB
0 0 0 SetCFg
1 1 1 SetCBg
0 SetP
[ 1 0 0 1 39 131 ] concat
143 353 185 379 Line
End

End 

Begin 
1 0 1 [] 0 SetB
0 0 0 SetCFg
1 1 1 SetCBg
0 SetP
[ 1 0 0 1 39 131 ] concat
406 348 372 298 Line
End

Begin 
1 0 1 [] 0 SetB
0 0 0 SetCFg
1 1 1 SetCBg
0 SetP
[ 1 0 0 1 39 131 ] concat
384 276 380 228 Line
End

Begin 
1 0 1 [] 0 SetB
0 0 0 SetCFg
1 1 1 SetCBg
0 SetP
[ 1 0 0 1 39 131 ] concat
236 319 322 347 Line
End

End 

showpage

end
 300 79 a
 endTexFig
143 2656 a Fa(Fig.)32 b(2)p eop
bop 117 165 a
 27102085 8683193 6183485 34601205 33285570 43284398 startTexFig
117 165 a

/IdrawDict 53 dict def
IdrawDict begin

/reencodeISO {
dup dup findfont dup length dict begin
{ 1 index /FID ne { def }{ pop pop } ifelse } forall
/Encoding ISOLatin1Encoding def
currentdict end definefont
} def

/ISOLatin1Encoding [
/.notdef/.notdef/.notdef/.notdef/.notdef/.notdef/.notdef/.notdef
/.notdef/.notdef/.notdef/.notdef/.notdef/.notdef/.notdef/.notdef
/.notdef/.notdef/.notdef/.notdef/.notdef/.notdef/.notdef/.notdef
/.notdef/.notdef/.notdef/.notdef/.notdef/.notdef/.notdef/.notdef
/space/exclam/quotedbl/numbersign/dollar/percent/ampersand/quoteright
/parenleft/parenright/asterisk/plus/comma/minus/period/slash
/zero/one/two/three/four/five/six/seven/eight/nine/colon/semicolon
/less/equal/greater/question/at/A/B/C/D/E/F/G/H/I/J/K/L/M/N
/O/P/Q/R/S/T/U/V/W/X/Y/Z/bracketleft/backslash/bracketright
/asciicircum/underscore/quoteleft/a/b/c/d/e/f/g/h/i/j/k/l/m
/n/o/p/q/r/s/t/u/v/w/x/y/z/braceleft/bar/braceright/asciitilde
/.notdef/.notdef/.notdef/.notdef/.notdef/.notdef/.notdef/.notdef
/.notdef/.notdef/.notdef/.notdef/.notdef/.notdef/.notdef/.notdef
/.notdef/dotlessi/grave/acute/circumflex/tilde/macron/breve
/dotaccent/dieresis/.notdef/ring/cedilla/.notdef/hungarumlaut
/ogonek/caron/space/exclamdown/cent/sterling/currency/yen/brokenbar
/section/dieresis/copyright/ordfeminine/guillemotleft/logicalnot
/hyphen/registered/macron/degree/plusminus/twosuperior/threesuperior
/acute/mu/paragraph/periodcentered/cedilla/onesuperior/ordmasculine
/guillemotright/onequarter/onehalf/threequarters/questiondown
/Agrave/Aacute/Acircumflex/Atilde/Adieresis/Aring/AE/Ccedilla
/Egrave/Eacute/Ecircumflex/Edieresis/Igrave/Iacute/Icircumflex
/Idieresis/Eth/Ntilde/Ograve/Oacute/Ocircumflex/Otilde/Odieresis
/multiply/Oslash/Ugrave/Uacute/Ucircumflex/Udieresis/Yacute
/Thorn/germandbls/agrave/aacute/acircumflex/atilde/adieresis
/aring/ae/ccedilla/egrave/eacute/ecircumflex/edieresis/igrave
/iacute/icircumflex/idieresis/eth/ntilde/ograve/oacute/ocircumflex
/otilde/odieresis/divide/oslash/ugrave/uacute/ucircumflex/udieresis
/yacute/thorn/ydieresis
] def
/Helvetica-Bold reencodeISO def
/Courier-Bold reencodeISO def
/Times-Bold reencodeISO def

/arrowHeight 8 def
/arrowWidth 4 def
/none null def
/numGraphicParameters 17 def
/stringLimit 65535 def

/Begin {
save
numGraphicParameters dict begin
} def

/End {
end
restore
} def

/SetB {
dup type /nulltype eq {
pop
false /brushRightArrow idef
false /brushLeftArrow idef
true /brushNone idef
} {
/brushDashOffset idef
/brushDashArray idef
0 ne /brushRightArrow idef
0 ne /brushLeftArrow idef
/brushWidth idef
false /brushNone idef
} ifelse
} def

/SetCFg {
/fgblue idef
/fggreen idef
/fgred idef
} def

/SetCBg {
/bgblue idef
/bggreen idef
/bgred idef
} def

/SetF {
/printSize idef
/printFont idef
} def

/SetP {
dup type /nulltype eq {
pop true /patternNone idef
} {
dup -1 eq {
/patternGrayLevel idef
/patternString idef
} {
/patternGrayLevel idef
} ifelse
false /patternNone idef
} ifelse
} def

/BSpl {
0 begin
storexyn
newpath
n 1 gt {
0 0 0 0 0 0 1 1 true subspline
n 2 gt {
0 0 0 0 1 1 2 2 false subspline
1 1 n 3 sub {
/i exch def
i 1 sub dup i dup i 1 add dup i 2 add dup false subspline
} for
n 3 sub dup n 2 sub dup n 1 sub dup 2 copy false subspline
} if
n 2 sub dup n 1 sub dup 2 copy 2 copy false subspline
patternNone not brushLeftArrow not brushRightArrow not and and { ifill } if
brushNone not { istroke } if
0 0 1 1 leftarrow
n 2 sub dup n 1 sub dup rightarrow
} if
end
} dup 0 4 dict put def

/Circ {
newpath
0 360 arc
patternNone not { ifill } if
brushNone not { istroke } if
} def

/CBSpl {
0 begin
dup 2 gt {
storexyn
newpath
n 1 sub dup 0 0 1 1 2 2 true subspline
1 1 n 3 sub {
/i exch def
i 1 sub dup i dup i 1 add dup i 2 add dup false subspline
} for
n 3 sub dup n 2 sub dup n 1 sub dup 0 0 false subspline
n 2 sub dup n 1 sub dup 0 0 1 1 false subspline
patternNone not { ifill } if
brushNone not { istroke } if
} {
Poly
} ifelse
end
} dup 0 4 dict put def

/Elli {
0 begin
newpath
4 2 roll
translate
scale
0 0 1 0 360 arc
patternNone not { ifill } if
brushNone not { istroke } if
end
} dup 0 1 dict put def

/Line {
0 begin
2 storexyn
newpath
x 0 get y 0 get moveto
x 1 get y 1 get lineto
brushNone not { istroke } if
0 0 1 1 leftarrow
0 0 1 1 rightarrow
end
} dup 0 4 dict put def

/MLine {
0 begin
storexyn
newpath
n 1 gt {
x 0 get y 0 get moveto
1 1 n 1 sub {
/i exch def
x i get y i get lineto
} for
patternNone not brushLeftArrow not brushRightArrow not and and { ifill } if
brushNone not { istroke } if
0 0 1 1 leftarrow
n 2 sub dup n 1 sub dup rightarrow
} if
end
} dup 0 4 dict put def

/Poly {
3 1 roll
newpath
moveto
-1 add
{ lineto } repeat
closepath
patternNone not { ifill } if
brushNone not { istroke } if
} def

/Rect {
0 begin
/t exch def
/r exch def
/b exch def
/l exch def
newpath
l b moveto
l t lineto
r t lineto
r b lineto
closepath
patternNone not { ifill } if
brushNone not { istroke } if
end
} dup 0 4 dict put def

/Text {
ishow
} def

/idef {
dup where { pop pop pop } { exch def } ifelse
} def

/ifill {
0 begin
gsave
patternGrayLevel -1 ne {
fgred bgred fgred sub patternGrayLevel mul add
fggreen bggreen fggreen sub patternGrayLevel mul add
fgblue bgblue fgblue sub patternGrayLevel mul add setrgbcolor
eofill
} {
eoclip
originalCTM setmatrix
pathbbox /t exch def /r exch def /b exch def /l exch def
/w r l sub ceiling cvi def
/h t b sub ceiling cvi def
/imageByteWidth w 8 div ceiling cvi def
/imageHeight h def
bgred bggreen bgblue setrgbcolor
eofill
fgred fggreen fgblue setrgbcolor
w 0 gt h 0 gt and {
l b translate w h scale
w h true [w 0 0 h neg 0 h] { patternproc } imagemask
} if
} ifelse
grestore
end
} dup 0 8 dict put def

/istroke {
gsave
brushDashOffset -1 eq {
[] 0 setdash
1 setgray
} {
brushDashArray brushDashOffset setdash
fgred fggreen fgblue setrgbcolor
} ifelse
brushWidth setlinewidth
originalCTM setmatrix
stroke
grestore
} def

/ishow {
0 begin
gsave
fgred fggreen fgblue setrgbcolor
printFont printSize scalefont setfont
/descender 0 printFont /FontBBox get 1 get printFont /FontMatrix
get transform exch pop def
/vertoffset 1 printSize sub descender sub def {
0 vertoffset moveto show
/vertoffset vertoffset printSize sub def
} forall
grestore
end
} dup 0 2 dict put def

/patternproc {
0 begin
/patternByteLength patternString length def
/patternHeight patternByteLength 8 mul sqrt cvi def
/patternWidth patternHeight def
/patternByteWidth patternWidth 8 idiv def
/imageByteMaxLength imageByteWidth imageHeight mul
stringLimit patternByteWidth sub min def
/imageMaxHeight imageByteMaxLength imageByteWidth idiv patternHeight idiv
patternHeight mul patternHeight max def
/imageHeight imageHeight imageMaxHeight sub store
/imageString imageByteWidth imageMaxHeight mul patternByteWidth add string def
0 1 imageMaxHeight 1 sub {
/y exch def
/patternRow y patternByteWidth mul patternByteLength mod def
/patternRowString patternString patternRow patternByteWidth getinterval def
/imageRow y imageByteWidth mul def
0 patternByteWidth imageByteWidth 1 sub {
/x exch def
imageString imageRow x add patternRowString putinterval
} for
} for
imageString
end
} dup 0 12 dict put def

/min {
dup 3 2 roll dup 4 3 roll lt { exch } if pop
} def

/max {
dup 3 2 roll dup 4 3 roll gt { exch } if pop
} def

/arrowhead {
0 begin
transform originalCTM itransform
/taily exch def
/tailx exch def
transform originalCTM itransform
/tipy exch def
/tipx exch def
/dy tipy taily sub def
/dx tipx tailx sub def
/angle dx 0 ne dy 0 ne or { dy dx atan } { 90 } ifelse def
gsave
originalCTM setmatrix
tipx tipy translate
angle rotate
newpath
0 0 moveto
arrowHeight neg arrowWidth 2 div lineto
arrowHeight neg arrowWidth 2 div neg lineto
closepath
patternNone not {
originalCTM setmatrix
/padtip arrowHeight 2 exp 0.25 arrowWidth 2 exp mul add sqrt brushWidth mul
arrowWidth div def
/padtail brushWidth 2 div def
tipx tipy translate
angle rotate
padtip 0 translate
arrowHeight padtip add padtail add arrowHeight div dup scale
arrowheadpath
ifill
} if
brushNone not {
originalCTM setmatrix
tipx tipy translate
angle rotate
arrowheadpath
istroke
} if
grestore
end
} dup 0 9 dict put def

/arrowheadpath {
newpath
0 0 moveto
arrowHeight neg arrowWidth 2 div lineto
arrowHeight neg arrowWidth 2 div neg lineto
closepath
} def

/leftarrow {
0 begin
y exch get /taily exch def
x exch get /tailx exch def
y exch get /tipy exch def
x exch get /tipx exch def
brushLeftArrow { tipx tipy tailx taily arrowhead } if
end
} dup 0 4 dict put def

/rightarrow {
0 begin
y exch get /tipy exch def
x exch get /tipx exch def
y exch get /taily exch def
x exch get /tailx exch def
brushRightArrow { tipx tipy tailx taily arrowhead } if
end
} dup 0 4 dict put def

/midpoint {
0 begin
/y1 exch def
/x1 exch def
/y0 exch def
/x0 exch def
x0 x1 add 2 div
y0 y1 add 2 div
end
} dup 0 4 dict put def

/thirdpoint {
0 begin
/y1 exch def
/x1 exch def
/y0 exch def
/x0 exch def
x0 2 mul x1 add 3 div
y0 2 mul y1 add 3 div
end
} dup 0 4 dict put def

/subspline {
0 begin
/movetoNeeded exch def
y exch get /y3 exch def
x exch get /x3 exch def
y exch get /y2 exch def
x exch get /x2 exch def
y exch get /y1 exch def
x exch get /x1 exch def
y exch get /y0 exch def
x exch get /x0 exch def
x1 y1 x2 y2 thirdpoint
/p1y exch def
/p1x exch def
x2 y2 x1 y1 thirdpoint
/p2y exch def
/p2x exch def
x1 y1 x0 y0 thirdpoint
p1x p1y midpoint
/p0y exch def
/p0x exch def
x2 y2 x3 y3 thirdpoint
p2x p2y midpoint
/p3y exch def
/p3x exch def
movetoNeeded { p0x p0y moveto } if
p1x p1y p2x p2y p3x p3y curveto
end
} dup 0 17 dict put def

/storexyn {
/n exch def
/y n array def
/x n array def
n 1 sub -1 0 {
/i exch def
y i 3 2 roll put
x i 3 2 roll put
} for
} def

Begin
[ 0.923077 0 0 0.923077 0 0 ] concat
/originalCTM matrix currentmatrix def

Begin 
1 0 0 [] 0 SetB
0 0 0 SetCFg
1 1 1 SetCBg
0.75 SetP
[ 1 0 0 1 90 192 ] concat
38 402 430 494 Rect
End

Begin 
1 0 0 [2 2 2 2 2 2 2 2] 15 SetB
0 0 0 SetCFg
1 1 1 SetCBg
0.75 SetP
[ 1 0 0 1 104 191 ] concat
123 496 123 403 Line
End

Begin 
1 0 0 [2 2 2 2 2 2 2 2] 15 SetB
0 0 0 SetCFg
1 1 1 SetCBg
0.75 SetP
[ 1 0 0 1 85 191 ] concat
238 495 238 404 Line
End

Begin 
1 0 0 [2 2 2 2 2 2 2 2] 15 SetB
0 0 0 SetCFg
1 1 1 SetCBg
0.75 SetP
[ 1 0 0 1 97 191 ] concat
324 494 324 404 Line
End

Begin 
0 0 0 SetCFg
Helvetica-Bold 14 SetF
[ 1 0 0 1 463 643 ] concat
[
(III)
] Text
End

Begin 
0 0 0 SetCFg
Helvetica-Bold 14 SetF
[ 1 0 0 1 362 642 ] concat
[
(II)
] Text
End

Begin 
0 0 0 SetCFg
Helvetica-Bold 14 SetF
[ 1 0 0 1 269 643 ] concat
[
(I)
] Text
End

Begin 
0 0 0 SetCFg
Helvetica-Bold 14 SetF
[ 1 0 0 1 149 645 ] concat
[
(Source)
] Text
End

Begin 
0 0 0 SetCFg
Helvetica-Bold 14 SetF
[ 0 -1 1 0 543 654 ] concat
[
(Sink)
] Text
End

Begin 
1 0 0 [] 0 SetB
0 0 0 SetCFg
1 1 1 SetCBg
0 SetP
[ 1 0 0 1 95 133 ] concat
427 554
431 548
2 BSpl
End

Begin 
1 0 0 [] 0 SetB
0 0 0 SetCFg
1 1 1 SetCBg
0 SetP
[ 1 0 0 1 95 133 ] concat
431 547
427 540
2 BSpl
End

Begin 
1 0 0 [] 0 SetB
0 0 0 SetCFg
1 1 1 SetCBg
0 SetP
[ 1 0 0 1 95 133 ] concat
427 540
431 530
2 BSpl
End

Begin 
1 0 0 [] 0 SetB
0 0 0 SetCFg
1 1 1 SetCBg
0 SetP
[ 1 0 0 1 95 133 ] concat
431 530
427 522
2 BSpl
End

Begin 
1 0 0 [] 0 SetB
0 0 0 SetCFg
1 1 1 SetCBg
0 SetP
[ 1 0 0 1 95 133 ] concat
425 522
431 508
2 BSpl
End

Begin 
1 0 0 [] 0 SetB
0 0 0 SetCFg
1 1 1 SetCBg
0 SetP
[ 1 0 0 1 95 133 ] concat
430 509
427 501
2 BSpl
End

Begin 
1 0 0 [] 0 SetB
0 0 0 SetCFg
1 1 1 SetCBg
0 SetP
[ 1 0 0 1 95 133 ] concat
427 501
430 488
2 BSpl
End

Begin 
1 0 0 [] 0 SetB
0 0 0 SetCFg
1 1 1 SetCBg
0 SetP
[ 1 0 0 1 95 133 ] concat
430 489
427 482
2 BSpl
End

Begin 
1 0 0 [] 0 SetB
0 0 0 SetCFg
1 1 1 SetCBg
0 SetP
[ 1 0 0 1 95 133 ] concat
427 483
431 471
2 BSpl
End

Begin 
1 0 0 [] 0 SetB
0 0 0 SetCFg
1 1 1 SetCBg
0 SetP
[ 1 0 0 1 95 133 ] concat
430 472
427 463
2 BSpl
End

Begin 
0 0 0 SetCFg
Courier-Bold 12 SetF
[ 0 1 -1 0 113.5 611.5 ] concat
[
(Wall)
] Text
End

Begin 
0 0 0 SetCFg
Courier-Bold 12 SetF
[ 1 0 0 1 335 586 ] concat
[
(Wall)
] Text
End

Begin 
0 0 0 SetCFg
Courier-Bold 12 SetF
[ 1 0 0 1 308 708 ] concat
[
(Wall)
] Text
End

Begin 
0 0 0 SetCFg
Times-Bold 14 SetF
[ 1 0 0 1 215 587 ] concat
[
(x)
] Text
End

Begin 
0 0 0 SetCFg
Times-Bold 14 SetF
[ 1 0 0 1 106 672 ] concat
[
(y)
] Text
End

Begin 
1 0 1 [] 0 SetB
0 0 0 SetCFg
1 1 1 SetCBg
0 SetP
[ 1 0 0 1 -166 133 ] concat
276 447 373 447 Line
End

Begin 
1 0 1 [] 0 SetB
0 0 0 SetCFg
1 1 1 SetCBg
0 SetP
[ 1 0 0 1 91 107 ] concat
19 474 19 544 Line
End

End 

showpage

end
 117 165 a
 endTexFig
143 2656 a Fa(Fig.)32 b(1)p eop
end
userdict /end-hook known{end-hook}if